# Tuning relaxation and nonlinear upconversion of valley-exciton-polaritons in a monolayer semiconductor

Hangyong Shan[1], Jamie M. Fitzgerald[2], Roberto Rosati[2], Gilbert Leibeling[4,5,6], Kenji Watanabe[8], Takashi Taniguchi[9], Seth Ariel Tongay[7], Falk Eilenberger[4,5,6], Martin Esmann[1], Sven Höfling[3], Ermin Malic[2] and Christian Schneider[1]

[1]*Institute of Physics, Carl von Ossietzky University Oldenburg, 26129 Oldenburg, Germany*
[2]*Department of Physics, Philipps-Universität Marburg, 35032 Marburg, Germany*
[3]*Julius-Maximilians-Universität Würzburg, Physikalisches Institut and Würzburg-Dresden Cluster of Excellence ct.qmat, Lehrstuhl für Technische Physik, Am Hubland, 97074 Würzburg, Germany*
[4]*Institute of Applied Physics, Abbe Center of Photonics, Friedrich Schiller University Jena, 07745 Jena, Germany*
[5]*Fraunhofer-Institute for Applied Optics and Precision Engineering IOF, 07745 Jena, Germany*
[6]*Max-Planck-School of Photonics, 07745 Jena, Germany*
[7]*School for Engineering of Matter, Transport, and Energy, Arizona State University, Tempe, Arizona 85287, USA*
[8]*Research Center for Electronic and Optical Materials, National Institute for Materials Science, 1-1 Namiki, Tsukuba 305-0044, Japan*
[9]*Research Center for Materials Nanoarchitectonics, National Institute for Materials Science, 1-1 Namiki, Tsukuba 305-0044, Japan*

**ABSTRACT**

Controlling exciton relaxation and energy conversion pathways via their coupling to photonic modes is a central task in cavity-mediated quantum materials research. In this context, the light-matter hybridization in optical cavities can lead to intriguing effects, such as modified carrier transport, enhancement of optical quantum yield, and control of chemical reaction pathways. Here, we investigate the impact of the strong light-matter coupling regime on energy conversion, both in relaxation and upconversion schemes, by utilizing a strongly charged $MoSe_2$ monolayer embedded in a spectrally tunable open-access cavity. We find that the charge carrier gas yields a significantly modified photoluminescence response of cavity exciton-polaritons, dominated by an intra-cavity like pump scheme. In addition, upconversion luminescence emerges from a population transfer from fermionic trions to bosonic exciton-polaritons. Due to the availability of multiple optical modes in the tunable open cavity, it seamlessly meets the cavity-enhanced double resonance condition required for an efficient upconversion. The latter can be actively tuned via the cavity length in-situ, displaying nonlinear scaling in intensity and fingerprints of the valley polarization. This suggests mechanisms that include both trion-trion Auger scattering and phonon absorption as its underlying microscopic origin.

## INTRODUCTION

Atomically thin transition metal dichalcogenides (TMDCs) have emerged as a highly interesting class of materials for opto-electronic and nanophotonic applications[1-4]. Due to the enormous binding energies between electrons and holes, the spectral response of monolayer TMDCs is dominated by correlated many-body excitations, even at elevated temperatures[5, 6]. Furthermore, the enhanced stability of excitons paves the way to study the fundamentals of bosonic physics in the solid state and the interactions in Bose-Fermi mixtures of excitons and electron gases[7-9].

TMDC monolayers have also evolved into one of the most versatile platforms to study the fundamentals of light-matter coupling[10-12], thanks to their giant dipolar coupling strength. This enables the emergence of exciton-polaritons supported by various types of optical microcavities, and under a wide range of temperature and particle density conditions. Such studies include the formation of polaritonic condensates from cryogenic[13-15] to room temperature[16-18], the emergence of moire exciton-polaritons and nonlinear dipolaritons in van der Waals heterostructures[19-21], and even the formation of Fermi-polaron-polaritons in high doping conditions[22]. While many studies have addressed effects of free or charged electron gases on the polaritonic relaxation[22-26], open questions regarding the interplay of excitons and trions in the strong coupling regime remain, for instance, the possibility of energy upconversion between trions and exciton-polaritons.

Photon upconversion occurs when light emission has an energy greater than that of the absorbed photons. This anti-Stokes photoluminescence (PL) is one of the fundamental principles behind laser cooling in solids, where phonons are absorbed to provide a gain in energy[27, 28]. Efficient upconversion PL requires a condition where both the excitation laser and the emission are resonant with electronic transitions[29]. TMDC monolayers are ideal for this requirement[30-32], since they possess various bound excitations with an energy spacing fortuitously matched with typical phonon energies. In addition, TMCD monolayers are direct bandgap semiconductors, and this may overcomes the limitation of the small absorption coefficient possessed by rare-earth doped glasses, which have been extensively applied in bioimaging and photovoltaics on the basis of upconversion[33]. It has been reported that upconversion efficiency can be modulated by plasmonic[34, 35] and dielectric[36] cavities. However, passively tunable cavities, lacking flexible adjustment in resonance energy, are insufficient for practical opto-electronic applications; thus, the realization of actively tunable upconversion is desired.

In this work, we scrutinize the impact of trions in a strongly charged $MoSe_2$ monolayer in a tunable optical cavity on energy relaxation and upconversion. We couple two longitudinal modes of the open cavity with electronic transitions of a $MoSe_2$ monolayer, observing a cavity-mediated double resonance condition for efficient upconversion: one mode weakly couples to trions, while the other is on resonance with excitons, yielding exciton-polaritons through strong coupling. Our results verify that the polariton luminescence is strongly dominated by the trionic state in the regular PL experiment. Furthermore, efficient and actively-tunable upconversion luminescence, emitted from exciton-polaritons, is observed. It demonstrates features of nonlinearity in intensity and valley-polarization with respect to the pump laser polarization, hinting at multiple mechanisms associated with trion Auger-like interaction and phonon absorption.

## RESULTS

Figure 1a shows schematic diagrams of the sample structure. We use an optical open cavity, in which the cavity length can be precisely adjusted in-situ, to study interactions among photons, excitons, and trions in a charged $MoSe_2$ monolayer. The cavity length scanning is performed by tuning the piezo positioner that moves the bottom distributed Bragg reflectors (DBRs) in Z direction, allowing for a fine scanning of the air-gap over ~0.7 μm while maintaining nanometer resolution. The open cavity setup has a notable advantage to mediate a double resonance condition required for efficient upconversion: it has a series of longitudinal modes whose free spectral range can be tuned via the cavity length to match electronic transitions. Thus, the open cavity is a versatile platform for upconversion research.

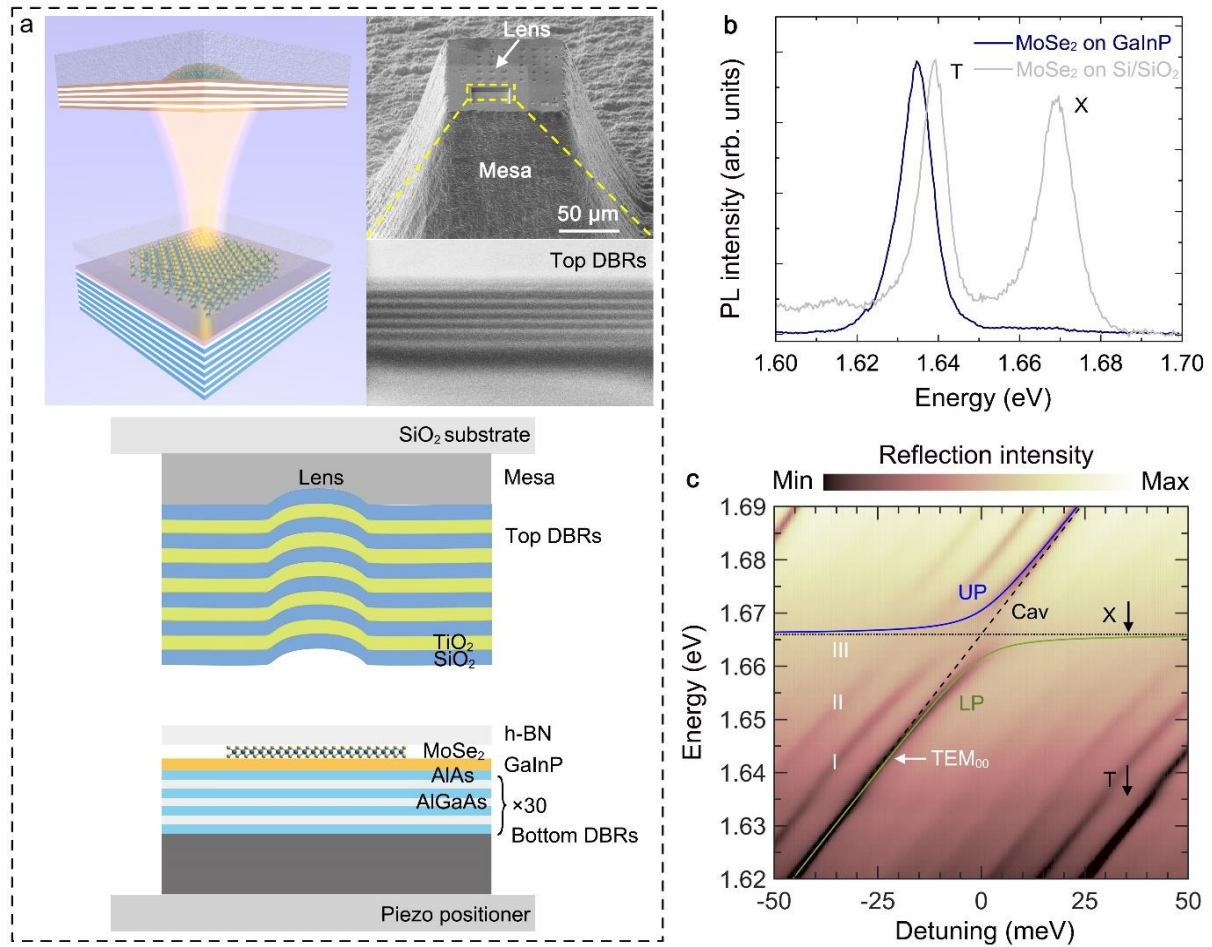

**Fig. 1 Sample sketch and strong coupling a** Schematic diagrams of the open cavity system and scanning electron microscope images of the top mirror. **b** PL of MoSe$_2$ monolayer on the bottom DBRs with a GaInP cap layer (without the top mirror), and PL of MoSe$_2$ on a conventional Si/SiO$_2$ substrate, at temperature of 3.5 K. **c** Reflection spectra recorded at different detuning. The anticrossing with MoSe$_2$ excitons (X) occurs at 1.666 eV, whereas the trions (T) only weakly perturb the optical resonance. Solid lines: fitting of a two-oscillator model between excitons and cavity photons of one longitudinal mode TEM$_{00}$.

The bottom DBRs are composed of 30 pairs of AlAs/AlGaAs layers, featuring a central Bragg wavelength of 750 nm. They are terminated by a 10 nm thick layer of GaInP, which facilitates free charge doping into the MoSe$_2$ monolayer[37]. The MoSe$_2$ monolayer is transferred on the GaInP layer and covered by a hexagonal-boron nitride (hBN) layer. The top mirror of the open cavity consists of 5.5 pairs SiO$_2$/TiO$_2$ DBRs. To further enhance the light-matter interaction, the in-plane mode confinement of the cavity is enabled by inclusion of a sphere cap-shaped lens. The entire open cavity system is loaded into a liquid helium-free optical cryostat (3.5 K) with ultra-low vibration and long-term stability. More details of the setup can be found in our previous work[38].

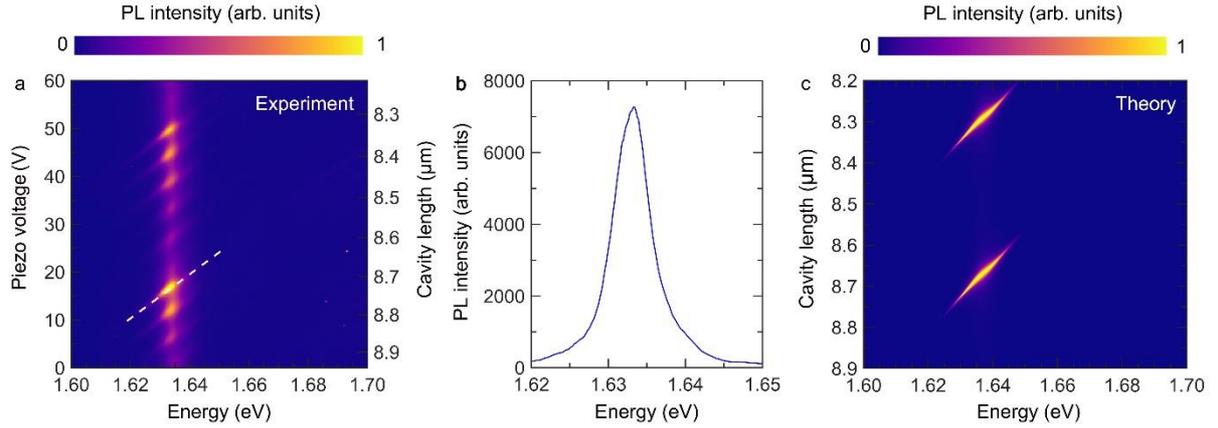

**Fig. 2 Open cavity modulated PL a** Experimentally measured PL of the sample with excitation of 532 nm laser. The emission is almost exclusively occurring as the detuned LPB (predominately photon-like) crosses the trion resonance. The right y-axis shows estimated cavity length at corresponding voltages. **b** Intensity profile of the PL as a function of energy. The profile path curve is indicated as the dashed line in panel **a**. **c** Theoretical model of PL intensity based on a single mode cavity (without the consideration of transverse modes), showing the weak coupling between the trion resonance and the strongly detuned LPB branch.

PL of the $MoSe_2$/GaInP/Bottom DBRs sample (without the top mirror) is plotted in Fig. 1b, together with the PL of a $MoSe_2$ monolayer on $Si/SiO_2$ substrate (grey curve) for comparison. The type II interface between $MoSe_2$ and GaInP, anticipated in a previous paper[37], yields an efficient carrier transfer from the GaInP layer to the monolayer, hence flooding the $MoSe_2$ with free charges and dramatically tipping the balance between neutral excitons and trions: compared to the reference sample prepared on a $Si/SiO_2$ substrate, the PL of the $MoSe_2$ monolayer on GaInP is trion-dominated, accompanied by an extremely subdued emission from excitons.

In contrast to previous reports on gated $MoSe_2$ monolayers[22], in our sample, the oscillator strength transfer from the exciton to an attractive polaron has not been accomplished, despite the evident substantial charging. As a consequence, in optical reflection spectra (Fig. 1c), when scanning the cavity length through the two resonances of the $MoSe_2$ monolayer, a strong coupling condition with a Rabi-splitting of 9 meV is notable only when the cavity approaches the exciton resonance. In contrast, when the cavity crosses the trion transition, it only exhibits weak coupling. The cavity mode that couples to the trion is a strongly detuned exciton-polariton resonance of primarily photonic character (~7% exciton fraction). Details of theoretical modeling based on the Hopfield approach can be found at Section 1 of the Supplementary Information (SI).

Due to the zero-dimensional confinement and the cylindrical symmetry of the hemispheric lens, in Fig. 1c we observe Laguerre-Gaussian-type transverse optical modes of the Fabry-Perot resonator in our open cavity[38-40], i.e., various transverse modes (labelled I, II, III, etc.) associated to one longitudinal mode ($TEM_{00}$).

Carrier relaxation in our system is studied via non-resonant injection (532 nm, CW pump) of electron-holes pairs, and detection of the steady state PL. In Fig. 2a, as a notable result, we only detect significant PL from the lower polariton branch (LPB) as its energy approaches the trion resonance. No PL is observed from exciton-polaritons around 1.666 eV. The left y-axis of the figure indicates the voltage supplied to the piezo positioner, while the right y-axis presents corresponding cavity lengths estimated via the free spectral range between two adjacent longitudinal modes. The spectral tuning curve along one longitudinal mode (dashed line in Fig. 2a) is plotted in Fig. 2b, where the intensity of the detuned polaritonic mode is displayed as a function of energy. This tuning curve qualitatively reproduces the PL spectrum of trions at low temperature (Fig. 1b), and suggests a direct energy conversion from trions to strongly detuned exciton-polaritons.

We model the emission of the coupled system via first calculating the absorption by using the transfer-matrix method. The PL can then be determined using the Kubo-Martin-Schwinger relation, which states that at thermodynamic equilibrium, the PL is proportional to the absorption scaled by the Boltzmann distribution[41] (details can be found in the Methods). Translational invariance is assumed in the in-plane direction, so transverse optical modes are neglected. The result of the model is shown in Fig. 2c and shows excellent agreement with the experiment.

In addition to the investigation of carrier relaxation from respect of regular PL experiment, the versatility of our open cavity setup allows us to study energy conversion pathways in more complex settings. Upconversion luminescence is an anti-Stokes process, in which the detected emission signal occurs at a higher energy than the optical pump. This process can arise from multiple microscopic origins, such as phonon absorption[29], which can yield effective laser cooling in solids[27], and excitation pair scattering into higher energy bands[42], as depicted in Fig. 3a.

In our study, we use energy-tunable cavity polariton modes to actively tune the upconversion process. We use 2 ps long excitation pulses with a central energy of 1.638 eV, i.e. spectrally resonant with the trion transition of $MoSe_2$. In Fig. 3b, as the cavity length is scanned, the emergence of upconversion luminescence occurs periodically whenever a resonance condition between the laser frequency/trion and the LPB is established. We see two groups of significant upconversion PL features, with the most intense emission at applied piezo voltages of 15.50 V and 51.25 V, respectively, featuring a double peak at energies of 1.659 eV and 1.665 eV. We note that the left edge of the lower energy peak with a maximum at 1.659 eV is partially cut by spectral filters that are used to separate the pump laser from upconversion signals.

The in-situ active tunability of the upconversion luminescence for a fine variation of the cavity length (piezo voltages around 15.50 V) is shown in Fig. S3 of the SI. This further supports our finding that resonance conditions are essential to observe strong upconversion PL. In addition, measurements with different excitation energies can be found in Fig S4, which verify that the laser spectral matching with the trion transition is also essential for a strong upconverted signal.

The peak energies of the upconverted PL match the polaritonic energies in Fig. 1c, and suggests their identification with the Rabi-split upper and lower polariton resonances. Specifically, we can assign these upconversion signals to exciton-polariton emission arising from higher energy transverse modes: At 15.50 V, the longitudinal mode $TEM_{00}$ matches the laser (and trion) energy, and efficient upconversion PL stems from polaritons, which consist of excitons strongly coupled to the transverse mode II. As the voltage is changed to 10.25 V, the pump is on resonance with the transverse mode I (Fig. S2), and we observe upconverted polariton emission, originating from a strong interaction between excitons and transverse mode III. The detailed mode structure analysis is provided in Fig. S2 of the SI.

Information about the microscopic origin of the upconversion luminescence is revealed by its power-dependence at various cavity-laser detunings. In Fig. 3c, we focus on the resonance at 15.50 V as an example. We show power-dependent upconversion PL, where the intensities of both peaks feature a nonlinear scaling with the pump power (see Fig 3d). Interestingly, the intensity of the lower energy peak gradually exceeds the high energy peak as the power increases. This phenomenon is attributed to a polariton bottleneck effect which is important at low densities, but can be overcome at higher densities (see further details in the Discussion).

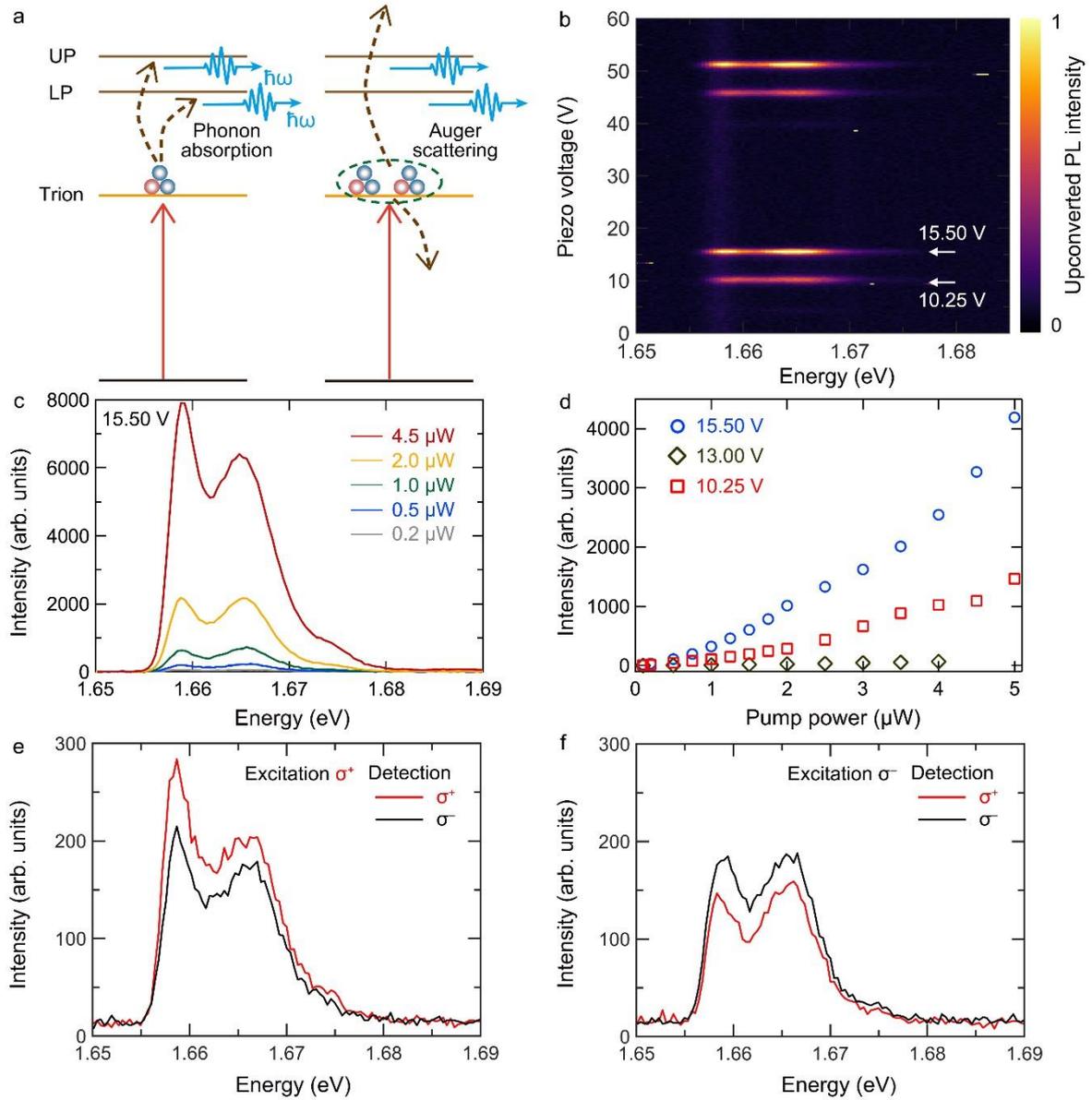

**Fig. 3 Nonlinear upconverted PL of valley-exciton-polaritons a** Diagrams of upconversion processes: phonon absorption and Auger scattering of trions that are mediated by cavity-like polaritons. **b** Upconversion PL plotted in false color scale as a function of piezo voltage (cavity length), with the excitation energy of 1.638 eV. **c** Pump power dependent upconversion intensity at a fixed piezo voltage of 15.50 V (resonance conditions between LPB and trion). **d** Integrated upconversion intensity as a function of pump power. The higher energy mode at 1.665 eV is analyzed. At 15.50 V and 10.25 V, intensive upconversion emission is observed, with a nonlinear increase versus pump power (nonlinear power law coefficient ~1.4). At 13.00 V, inefficient upconversion occurs due to being off-resonance. **e** Circular polarization of upconversion emission with σ⁺ excitation, **f** with σ⁻ excitation.

The power-dependent integrated intensity of the 1.665 eV peak, recorded at various piezo voltages, is shown in Fig. 3d. It reveals a nonlinear dependence of upconverted signal on pump power for different voltages. To extract the characteristic power-law coefficient, the log-log plot of the data from Fig. 3d is shown in Fig. S5. At 15.50 V and 10.25 V, power-law coefficients of 1.4 are extracted for both voltages. At the voltage 13.00 V, where the cavity-like LPB mode is detuned from the pump laser, we observe a reduced upconversion signal with a slightly smaller power law coefficient of 1.3 (Fig. S5).

It is insightful to compare the power-law coefficients of regular and upconversion PL. As shown in Fig. S1, we see a linear increase of luminescence intensity in regular PL, i.e., with a coefficient of 1, which contrasts with the coefficient of 1.4 for the upconversion PL. If upconverted luminescence solely originates from Auger-type scattering, the power-law coefficient can be expected to be approximately twice as large as that of regular PL[42], however, this condition does not fit our results. In contrast, phonon-assisted upconversion, where a trion absorbs one or multiple phonon(s) and dissociates into an exciton and free electron/hole, gives rise to a linear response[29]. Therefore, the intermediate power-law coefficients suggest the upconversion process involves both Auger-type scattering and direct phonon absorption.

A further fingerprint of the phonon-assisted upconversion process is revealed by polarization-resolved studies. In Figs. 3e,f, we show circular-polarization dependent upconversion PL. We find that upconverted exciton-polaritons retain some part of the pump laser polarization. In Ref.[29], phonon-assisted upconversion of exciton emission in $WSe_2$ monolayers was reported to partially preserve the valley polarization. Whereas regarding Auger scattering, strong valley depolarization is expected to occur during the relaxation from high energy states. Due to Rashba-induced coupling of the dark and bright exciton branches[43], the valley depolarization observed in $MoSe_2$ monolayers is much stronger than that in its $MoS_2$ and $WSe_2$ counterparts, leading to an intrinsic degree of circular polarization 10 times lower[44].

**DISCUSSION**

At first sight, phonon absorption seems an unlikely mechanism for the observed upconverted PL in our system since there will be a vanishingly small phonon population at the experimental temperature of 3.5 K[45]. Remarkably though, the observed polarization-retention and power law strongly suggest an important contribution of phonon-assisted upconversion, which is in agreement with prior cavity-free studies of $MoSe_2$[30, 46]. Thus, we propose a combination of physical mechanisms that lead to a significant phonon-driven upconverted signal from the polariton states.

First, the coincidental energy matching between the $A_1$ optical phonon mode and the exciton-trion splitting leads to a doubly resonant Raman scattering[46]. The small Rabi splitting of 9 meV preserves this near coincidence in energy between the trion and exciton-polaritons. Furthermore, the small number of phonons available for absorption can be partially compensated by a large trion population. Under the resonance excitation condition, efficient absorption of the $A_1$ phonon will take place and lead to a transient over-population of the exciton-polariton states, i.e., a much larger occupation than predicted by the Boltzmann ratio.

It is also important to consider the formation mechanism of the trion. At higher pump powers, the law of mass action[47] predicts a depletion of free electron population, which favours the formation of exciton-polaritons, resulting in a stronger upconversion signal at steady state. For further details, see Section 2 of the SI.

Furthermore, Auger scattering between trions leads to the decay of one trion and the excitation of a higher energetic state. This state will then relax toward the ground state by emitting a cascade of optical phonons[48]. This generation of hot phonons, which is dictated by the ratio of the trion energy to the phonon energies, heats the system and acts as a source for phonon-assisted upconversion.

This is supported by the measured pump-power dependent relative peak intensity of the upconverted signal in Fig 3c. In the case of a thermalized system, we would expect the lower-energy peak to dominate in intensity, but this is only observed at high pump power. We attribute this to a polariton bottleneck effect[49]: excited hot excitons get trapped in the high-energy (and momenta) exciton state and cannot relax efficiently because the energy jump between the states is small compared to a typical phonon energy of ~30 meV[46]. At higher pump power, the Auger mechanism yields transient excitation of very high-energy states, which would thermalize via multiple scattering events[48], hence increasing the opportunity to bypass the bottleneck. Additionally, a greater effective temperature of the system due to a hot phonon population could further reduce the bottleneck effect via increased phonon absorption[50].

In summary, the utilization of a GaInP layer achieves a substantial charge accumulation in MoSe$_2$ monolayers, resulting in trion-dominated PL, both, in the cavity-free case as well as the strong coupling regime. Efficient upconversion luminescence from exciton-polaritons can be observed, consistent with the condition that one detuned polariton mode is on resonance with the pump laser. The upconverted PL intensity exhibits a nonlinear dependence on pump power, with a power-law suggesting contributions from both trion-trion Auger scattering and phonon absorption. The preservation of valley polarization further confirms the role of phonon absorption, and provides an additional tuning knob for optoelectronic applications based on upconversion.

Our demonstration of upconversion from an intrinsic material-based resonance (trions) to hybrid quasi-particles (exciton-polaritons) outlines novel schemes for injection of bosonic polaritons from fermi reservoirs. The nonlinear behaviour reveals that upconversion can become an efficient injection method of polariton populations, and provides new insights into investigations of polariton condensation and many-body correlation physics of Bose-Fermi mixtures.

## METHODS

### Sample fabrication

The bottom DBRs are grown by molecular beam epitaxy, and consist of 30 pairs of AlAs/AlGaAs layers (61.5 nm/56.5 nm), and has a central wavelength of 750 nm. It is terminated by a 10 nm GaInP layer. On top of the DBRs, we assemble a MoSe$_2$ monolayer, capped with a multilayer hBN via the deterministic dry-transfer method. The top mirror of the open cavity is a SiO$_2$ mesa coated with 5.5 pairs of SiO$_2$/TiO$_2$ (130.0 nm/81.9 nm) DBRs via sputter evaporation. The SiO$_2$ mesa has a size of 100 μm x 100 μm, where sphere cap-shaped lenses with various diameters are milled with focused ion beam lithography. The utilized lens has a diameter of 6 μm and a depth of ~330 nm.

### Experimental setup

In regular PL measurements, the sample is excited with a 532 nm CW laser, while pulsed laser (Coherent Mira Optima 900-F mode-locked Ti:Sapphire laser) with 2 ps pulse duration is used for upconversion PL experiments. To filter out the excitation laser, we use 600 nm long-pass filter for regular PL, and 750 nm short-pass filters for upconversion measurements, respectively. In circular polarization measurements, the polarization control at the excitation and detection is calibrated with a polarization analyser (Schäfter+Kirchhoff SK010PA). In white-light reflection experiments, a 200 μm pinhole is mounted at the focal plane in real space to realize spatial filtering. All spectra are recorded using a spectrometer (Andor Shamrock SR-500i) attached with a CCD camera (Andor iKon-M934 Series). The open cavity is placed in a liquid helium-free optical cryostat (Attodry 1000), which features an ultra-low vibration, long-term stable measurement system. The whole setup was introduced in detail in our previous work[38].

### Theoretical description

Using the Kubo-Martin-Schwinger relation[41], the PL is obtained using a transfer-matrix calculation of the absorption, which is then scaled by the Boltzmann distribution evaluated at the experimental temperature of 3.5 K. Both the 1s exciton and trion energy are set to the experimental values. The exciton oscillator strength is chosen to match the experimental Rabi splitting of 9 meV. The oscillator strength and linewidth of the trion are extracted from measurements of a bare flake on the bottom DBRs. The ratio of the exciton and trion radiative decay rates is found to be ~130. We include a Gaussian broadening of 4 meV to mimic inhomogeneous broadening, and assume translational invariance in the in-plane direction, so transverse optical modes are neglected.

## DATA AVAILABILITY

The data that support the findings of this study are available from the corresponding authors upon reasonable request.

**Acknowledgement**

The authors acknowledge support by the German research foundation (DFG) via the project SCHN1376 14-2, funded within the priority program 2244. Financial support by the Niedersächsisches Ministerium für Wissenschaft und Kultur ("DyNano") is gratefully acknowledged. The Marburg group acknowledges funding from the DFG via SFB 1083 and the regular project 524612380. S.T. acknowledges support from U.S department of Energy DE-SC0020653. K.W. and T.T. acknowledge support from the JSPS KAKENHI (Grant Numbers 21H05233 and 23H02052) and World Premier International Research Center Initiative (WPI), MEXT, Japan. We thank Raül Perea-Causín (Stockholm University) for valuable discussions.

**COMPETING INTERESTS**

The authors declare no competing interests.